# Scout-Dose-TCM: Direct and Prospective Scout-Based Estimation of Personalized Organ Doses from Tube Current Modulated CT Exams

Maria Jose Medrano[1], Sen Wang[1], Liyan Sun[1], Abdullah-Al-Zubaer Imran[2], Jennie Cao[1], Grant Stevens[3], Justin Ruey Tse[1], Adam S. Wang[1]

[1]Stanford University, Department of Radiology, Stanford, CA 94305, USA

[2]University of Kentucky, Department of Computer Science, Lexington, KY 40506, USA

[3]GE HealthCare, Menlo Park, CA 94025, USA

**Objective:** This study proposes Scout-Dose-TCM for direct, prospective estimation of organ-level doses under tube current modulation (TCM) and compares its performance to two established methods.

**Approach:** We analyzed contrast-enhanced CT examinations of the chest-abdomen-pelvis among 130 adult patients with 120 kVp and TCM. Reference organ doses for six organs (lungs, kidneys, liver, pancreas, bladder, spleen) were retrospectively calculated using open-source tools MC-GPU and TotalSegmentator. Based on these reference doses, we trained Scout-Dose-TCM, a deep learning-based method that predicts organ-level doses corresponding to discrete cosine transform (DCT) basis functions to enable real-time estimation of organ doses under any TCM profile. The Scout-Dose-TCM model includes a feature learning module that extracts contextual information from lateral and frontal scouts and scan range, and a dose learning module that outputs dose basis estimates following the DCT formulation of TCM. A customized loss function was designed to incorporate the DCT formulation during training to ensure accurate dose prediction across variable TCM patterns. Two established methods were implemented for comparison: size-specific dose estimation following AAPM Task Group 204 (Global $CTDI_{vol}$) and its version adapted for TCM and organ-specific estimates (Organ $CTDI_{vol}$). A 5-fold cross-validation assessed generalizability by comparing mean absolute percentage dose errors and $R^2$ correlations with benchmark doses across the six studied organs.

**Main Results:** Overall absolute percentage errors for the six organs were 13% (Global $CTDI_{vol}$), 9% (Organ $CTDI_{vol}$), and 7% (Scout-Dose-TCM). The largest discrepancies were observed for the bladder, with errors of 15%, 13%, and 9%, respectively. Statistical analysis showed Scout-Dose-TCM significantly reduced organ-level dose errors compared to Global $CTDI_{vol}$ across most organs ($p < 1\times10^{-7}$) and provided notable improvements over Organ $CTDI_{vol}$ for the liver, bladder, and pancreas ($p \leq 0.005$). Scout-Dose-TCM also consistently achieved higher $R^2$ values, indicating stronger agreement with benchmark Monte Carlo estimates.

**Significance:** Our updated Scout-Dose-TCM model was more accurate than Global $CTDI_{vol}$ and comparable to better than Organ $CTDI_{vol}$, without requiring organ segmentations during inference. These results highlight the potential of Scout-Dose-TCM as a clinically useful tool for prospective organ-level dose estimation in CT.

## 1. INTRODUCTION

CT accounts for less than a quarter of all imaging procedures but are responsible for nearly half or more of the cumulative population radiation dose from medical imaging in the United States (Mettler, Thomadsen, *et al.*, 2008; US Food and Drug Administration, 2010; Smith-Bindman *et al.*, 2025). The use of CT continues to rise significantly and is expected to account for 5% of cancers if current radiation doses are maintained. As a result, in some states, radiologists are mandated to report radiation dose from CT and adhere to the as low as reasonably achievable ("ALARA") principle while maintaining diagnostic image quality (Mettler, Huda, *et al.*, 2008; Neumann and Bluemke, 2010; Thurston, 2010). To apply the ALARA principle in practice, radiologists and technologists require real-time estimates of a patient's potential radiation exposure before planning a CT. The most common first-order estimates of a patient's potential CT dose exposure can be reported as volume CT dose index ($CTDI_{vol}$) and dose-length product (DLP) values, which reflect the radiation output per unit scan length and the cumulative radiation output across the entire scan, respectively. However, these values are a measure of scanner output and do not directly represent patient-specific organ dose.

To better account for patient size, AAPM Task Group 204 introduced the size-specific dose estimate (SSDE) metric, which adjusts $CTDI_{vol}$ based on water-equivalent diameter (Boone *et al.*, 2011). While SSDE marked an important step forward, it does not capture dose variations introduced by tube current modulation (TCM), a common scanning technique that dynamically adjusts tube current in three dimensions to reduce radiation exposure while maintaining image quality (Kalender W.A. *et al.*, 1999; Kalender, Wolf and Suess, 1999; Greess *et al.*, 2000; McCollough, Bruesewitz and Kofler, 2006). For TCM scans, $CTDI_{vol}$ is calculated using the average tube current across the scan, yielding a single, non-specific estimate of patient dose. This fails to capture how dose is distributed across individual organs, since TCM dynamically adjusts tube current based on anatomical location and attenuation, leading to substantial organ-to-organ dose variations. Accounting for these differences is critical, as sensitivity to radiation dose varies amongst organs.

Some investigators have explored TCM optimization strategies that incorporate effective dose or patient-specific organ dose estimates (Klein *et al.*, 2022; Wang *et al.*, 2024). However, because these approaches rely on retrospective data to estimate patient-specific organ doses, they do not support prospective dose planning. Meanwhile, vendor-implemented organ-based TCM features remain limited: they typically apply only to select organs, rely on manual organ identification, and follow predefined rules for tube current reduction at specific angles (for example, anterior positions for the eye lens or breasts), without fully accounting for patient-specific organ doses (Duan *et al.*, 2011; Wang *et al.*, 2012; Gandhi *et al.*, 2015; Fillon *et al.*, 2018). As a result, only limited methods exist to prospectively adjust organ-level doses in TCM scans, constraining efforts to personalize risk assessment and imaging protocols. Advancing methods that can prospectively estimate organ-level dose under TCM is essential to bridge this gap.

Monte Carlo (MC)-based dose calculations provide patient-specific dose maps that account for attenuation and scan parameters, including TCM. However, they remain challenging to incorporate clinically due to high computational demands and their reliance on retrospectively generated 3D attenuation maps. To mitigate these issues, researchers have accelerated MC-based methods through parallelized X-ray photon transport simulations and GPU-based deterministic solvers for the linear Boltzmann transport equation (Badal and Badano, 2009; Wang et al., 2019; Principi et al., 2020). More recently, advances in deep learning have enabled direct prediction of MC-like dose distributions from CT images (Lee *et al.*, 2019; Fan *et al.*, 2020; Götz *et al.*, 2020). For instance, Maier et al. employed a 3D U-Net to replicate MC-calculated dose maps (Maier *et al.*, 2022), while Offe et al. combined an LBTE solver with V-Net segmentation for rapid pediatric CT dose assessment (Offe *et al.*, 2020). Other studies have applied computer vision to prospectively estimate dose by reconstructing 3D attenuation maps from scout views and performing automatic organ segmentation (Shen, Zhao and Xing, 2019; Montoya *et al.*, 2022). Nevertheless, these multi-step pipelines often introduce substantial computational overhead, and their accuracy may be constrained by errors in the intermediate 3D representations (Ying *et al.*, 2019; Almeida, Astudillo and Vandermeulen, 2021).

Alternative efforts have focused on calibration and empirical fitting methods that relate patient size, attenuation, and dose. Tian et al. used a virtual phantom library paired with a convolution-based dose-field model to estimate organ doses both prospectively and retrospectively (Tian *et al.*, 2016), while Gao et al. employed patient-matching approaches to predict organ and effective doses across large cohorts (Gao *et al.*, 2020). Moore et al. derived organ dose correlation factors from phantom studies for pediatric CT applications (Moore *et al.*, 2014), and Khatonabadi et al. refined the AAPM TG 204 framework to better correlate patient size metrics with organ doses (Khatonabadi *et al.*, 2013). Though informative, these approaches often rely on limited calibration phantoms that may not generalize to diverse anatomies or require organ segmentations, complicating prospective implementation. More recently, deep learning models have been proposed to directly estimate organ doses from scan parameters: Tzanis et al. incorporated slice-wise water-equivalent diameter, tube current, and HU statistics (Tzanis and Damilakis, 2024), while Myronakis et al. leveraged scan length, average water-equivalent diameter, and tube current to predict doses (Myronakis, Stratakis and Damilakis, 2023). However, these methods frequently depend on slice-level inputs or organ segmentations not available before scanning, or have been validated only under uniform tube current, limiting their applicability to real-time organ-level dose prediction in non-uniform TCM exams.

Building on the potential of deep-learning to extract encoded anatomical and attenuation information, our group previously developed the Scout-Net model: a fully automated, end-to-end CNN-based framework for real-time, patient-specific organ dose estimation from scout images (Imran *et al.*, 2021, 2022, 2023). Our initial work showed that organ doses could be prospectively predicted in real time using lateral and frontal scout images. However, the original Scout-Net model did not account for variations in organ dose introduced by TCM, a commonly used scanning mode in clinical practice. Furthermore, other approaches using deep-learning or physics models were limited by computational complexity, retrospective data requirements, or lack of integration with TCM. To address these limitations, we propose Scout-Dose-TCM, an enhanced model that integrates the periodic patterns of TCM profiles into the learning process to enable prospective, real-time, organ-level dose estimation for tube-current-modulated exams. By incorporating anatomical and attenuation information from scout images and the physics of TCM, Scout-Dose-

TCM estimates regional variations in organ dose that arise from TCM protocols. In this work, we also compare Scout-Dose-TCM against two established methods for prospective organ-level dose estimation, demonstrating the benefits of combining data-driven feature learning with physics-based dose modeling.

## 2. METHODS AND MATERIALS

Our proposed Scout-Dose-TCM allows for organ-level dose estimation under varying TCM by leveraging the periodic nature of TCM maps. To determine the potential impact of our proposed model, we compared it to two established prospective dose estimation methods in the literature: the original AAPM Task Group 204 (Boone *et al.*, 2011) and its updated version by Khatonabadi et al. for organ-level dose prediction with non-uniform TCM (Khatonabadi *et al.*, 2013). The following sections provide a detailed description of our Scout-Dose-TCM model implementation and comparative study. Section 2.1 outlines the generation of the benchmark retrospective organ-level dose values via Monte Carlo simulation. Section 2.2 describes the representation of our system's TCM tables which constitutes the basis of our updated Scout-Dose-TCM model. Section 2.3 presents two existing methods for prospective organ-level dose estimation, followed by our proposed Scout-Dose-TCM model. Finally, Section 2.4 describes the performance evaluation conducted to compare the three prospective dose estimation methods in our study.

*2.1. Retrospective Calculation of Reference Organ Doses*

*Monte Carlo Simulation Code and Source Model*

The retrospective organ-level dose benchmark values in our study were generated using an open-source Monte Carlo software, MC-GPU (Badal and Badano, 2009), modified to model a multidetector row helical CT scanner (Revolution CT, GE HealthCare, Waukesha, WI). The final MC-GPU simulation was designed to model our system's helical CT trajectory with polyenergetic spectra and to incorporate scanner-specific parameters including bowtie filter, anode heel effect, and tube current modulation. The system's 'Body' bowtie filter was simulated using the vendor-specific material thicknesses and corresponding attenuation coefficients to account for filtered spectra and relative photon counts at all fan angles. Scattered photons from the bowtie filter were ignored given their small contribution to the patient dose. The system's anode heel effect was modeled as a probability function to represent variation in x-ray intensity in the cone angle (azimuthal angle θ) following vendor specifications. To incorporate non-uniform tube current modulation, we assumed that the total photon flux per projection is proportional to the tube current, given an equal exposure time across all views. MC-GPU was modified to score the deposited energy and radiation dose resulting from interactions with the voxelized phantom at each projection, and organ doses for a specific TCM map were calculated as the weighted sum of single-view dose distributions. The accuracy of our simulation was validated by comparing CTDI results from phantom simulation in our MC-GPU framework against experimental measurements, yielding final validation errors within 1.5% (Wang S *et al.*, 2021; Wang *et al.*, 2025).

Our final simulations were run in helical mode (pitch factor: 0.99, detector z coverage: 80 mm) with a 120 kV source spectrum. MC-GPU dose simulations were repeated with $\Theta$ uniformly spaced start angles ($\Theta = 4$, $\delta(i) = \{0, 90, 180, 270\}°$). Considering that the actual start angles cannot be controlled prospectively, we averaged the $\Theta$ dose maps to obtain the dose map (Sharma *et al.*, 2019).

$$D_{avg} = \frac{1}{\Theta} \sum_{i=1}^{\Theta} MCGPU(V, \Psi)|_{\delta(i)} \qquad \text{Eq. 1}$$

where $V$ is the voxelized phantom and $\Psi$ denotes the set of all the parameters used to configure the MC-GPU simulation. MC-GPU doses were reported in eV/g/photon and scaled to the more standard mGy based on an empirical conversion factor from CTDI measurements. The computational times of our simulations differed based on z coverage of each patient scan, generally taking 1-2 hours to perform projection-based dose calculations for four different starting angles on a workstation with 2 NVIDIA TITAN RTXs and Intel Core i9-9960X CPU @ 3.10 GHz. Since dose is proportional to tube current, scaling the tube current modulation to a relative dose allowed for the estimation of organ-level doses using the same scaling. This approach eliminated the need for additional Monte Carlo simulations. Table I lists the modified set of parameters for our MC-GPU configuration.

**Table I:** Monte Carlo simulation parameters used in this study. Default MC-GPU parameters are excluded.

| Simulation Parameters | Values |
| --- | --- |
| Tube potential | 120 kVp |
| Z-axis coverage | 80 mm |
| Pitch factor | 0.99 |
| Voxel spacings (isotropic) | (4, 4, 4) mm$^3$ |
| Source to rotation axis distance | 625.61 mm |
| Source to detector distance | 1097.61 mm |
| Vertical translation between projections | 3.3 mm |
| Polar and azimuthal apertures | (42.0°, 7.32°) |
| Angle between projections | 15° |
| Views per rotation | 24 |
| Start angles | 0°, 90°, 180°, 270° |

*Patient Data and Benchmark Organ-level Dose Calculations*

CT of 130 adult patients undergoing outpatient contrast-enhanced chest-abdomen-pelvis examinations in single-energy mode at 120 kVp were exported from a GEHC Revolution scanner at Stanford Health Care under IRB approval with a waiver of informed consent. Corresponding scout images and TCM maps were extracted to implement retrospective and prospective organ-level dose estimations. Patient-specific voxelized phantoms with spatial maps of material type and mass density were derived from each patient CT image by following a standard piecewise linear curve defining the densities of the mixture of water and bone (Sharma *et al.*, 2019). Original CT images with dimensions 0.98 × 0.98 × 2.5 mm$^3$ were resampled to isotropic voxels of 4 × 4 × 4 mm$^3$ to ensure compatibility with our available computational resources. Segmentations for the 6 main organs of interest (lungs, kidneys, liver, bladder, spleen, and pancreas) were derived for each patient CT scan with the open-source segmentation tool TotalSegmentator (Wasserthal *et al.*, 2023). Final organ segmentation masks were used to extract organ-level dose values from the dose maps estimated with our customized MC-GPU simulations. Figure 1 provides a visual overview of the process followed to generate benchmark organ-level doses in our study.

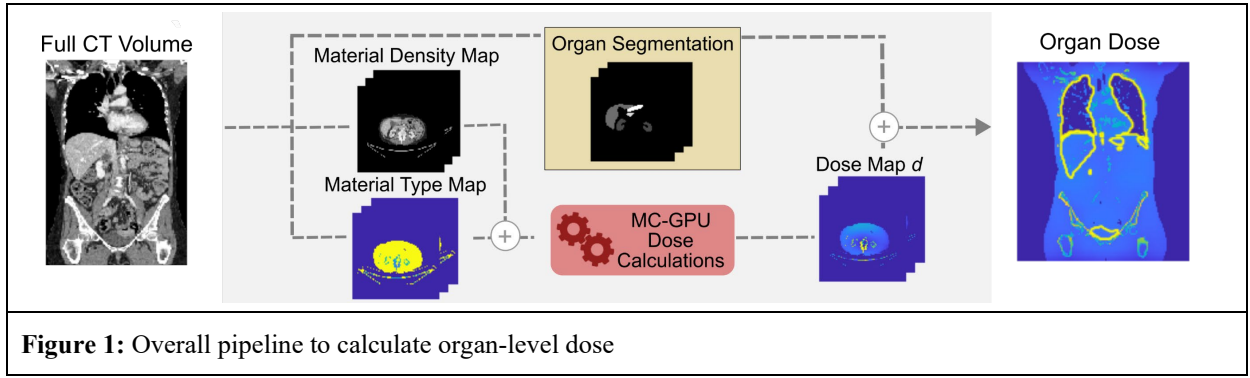

**Figure 1:** Overall pipeline to calculate organ-level dose

*2.2 Representation of Tube Current Modulation Maps with Discrete Cosine Transform*

Patient-specific TCM maps were generated using the vendor's proprietary TCM algorithm. We collected the tube current profile for each patient scan, which contains information on the tube current at each gantry position including gantry angle ($\phi$ from 0 to 360°, shown left to right) and bed position ($z$), enabling the tracing of a TCM profile for any arbitrary starting angle in a helical CT acquisition. Building on previous work (Wang *et al.*, 2024), each TCM map is expressed as a linear combination of $N$ discrete cosine transform (DCT) basis functions ($B_n$) and their corresponding coefficients $\beta_n$ to build a 2D TCM map that is periodic in the horizontal (gantry angle) direction. Figure 2 illustrates the TCM map and its DCT basis representation. Given the linear nature of this representation, the final organ-level dose ($D_{k,l}$) for organ $l$ under TCM map $k$ can be computed as a linear combination of basis doses ($d_{n,l}$) weighted by the DCT coefficient ($\beta_{k,n}$) as given by,

$$D_{k,l} = \sum_{n=1}^{N} \beta_{k,n} d_{n,l} \qquad \text{Eq. 2}$$

where $\beta_{k,n}$ is the $n$-th DCT coefficient for TCM map $k$, and $d_{n,l}$ is the basis dose of organ $l$ under DCT basis $B_n$. For this study, each TCM map was represented using N=27 DCT basis functions which led to a residual representation error of <5% in the TCM maps.

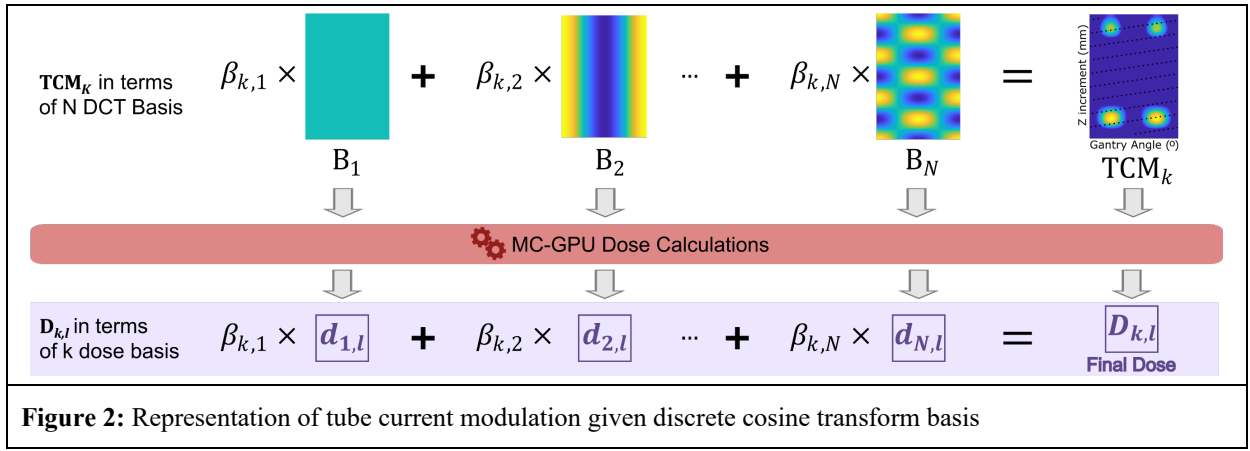

**Figure 2:** Representation of tube current modulation given discrete cosine transform basis

## 2.3. Prospective Estimation of Organ-Level Doses

*Proposed Scout-Dose-TCM Method*

Our previously proposed Scout-Net model leveraged deep-learning to estimate mean organ-level doses from scout images under uniform TCM maps (Imran *et al.*, 2021, 2022, 2023). However, in clinical practice most CT scans are acquired with patient-specific TCM to reduce patient radiation dose while maintaining diagnostic image quality (Greess *et al.*, 2000; McCollough, Bruesewitz and Kofler, 2006). To enable Scout-Dose-TCM to be clinically applicable, we trained our model to generalize not only to different patient attenuations but also an infinite combination of TCM maps by leveraging the DCT representation of TCM maps (Section 2.2). Instead of estimating the organ-level doses for a particular TCM ($D_{k,l}$), our Scout-Net model was updated to estimate the organ-level doses ($d_{n,l}$) for a set of $N$ discrete cosine basis functions ($B_n$). In doing so, we enable Scout-Dose-TCM to estimate the final dose $D_{k,l}$ for any TCM with discrete cosine coefficients $\beta_{k,n}$ following Eq. 2.

Our Scout-Dose-TCM model architecture consists of two modules: a feature learning module (FLM) and dose learning modules (DLM). The feature learning module extracts anatomical and attenuation details from scout images, and the dose learning modules infer the relationship between extracted features and organ-level doses. The three inputs of our model consist of frontal scouts ($x_{fs}$), lateral scouts ($x_{ls}$), and the patient scan range ($x_{sr}$) to predict the organ-level doses $d_{l,n}$ for $N = 27$ discrete cosine bases and $L = 6$ organs. While frontal and lateral scouts provide information on patient attenuation and anatomy, scan range was incorporated as an input to inform our model of the scanned region in the scouts. Frontal and lateral scouts were registered and aligned to the scan range, which were prepared as 750 × 530 pixel (1 × 1 mm² per pixel) images based on the full CT scan range. The $x_{fs}$, $x_{ls}$, and $x_{sr}$ were then concatenated to make a 3-channel input to our Scout-Dose-TCM model of size 750 × 530 × 3. Our FLM consists of a convolutional backbone comprising four sequential blocks, each containing a 3×3 convolutional layer with padding, instance normalization, and a LeakyReLU activation function with a negative slope of 0.2. Spatial downsampling (stride = 2) was applied in the second and fourth blocks. A global average pooling layer was used to aggregate spatial features into a 128-dimensional representation. This was followed by the model's dose learning module (DLM), which consists of two fully connected layers: a 64-unit layer followed by a final linear layer that produces a 27-dimensional output representing normalized dose estimates ranging from 0 to 1. For this implementation, separate models were trained for each organ of interest to allow for organ-specific feature learning and improved prediction accuracy, and a joint loss function was computed across all organ-specific models during training to encourage coordinated optimization and balanced performance. This design allowed targeted optimization and interpretability per organ, which was beneficial given the distinct imaging and dose characteristics of each anatomical region. The final Scout-Net-TCM model architecture is illustrated in Figure 3.

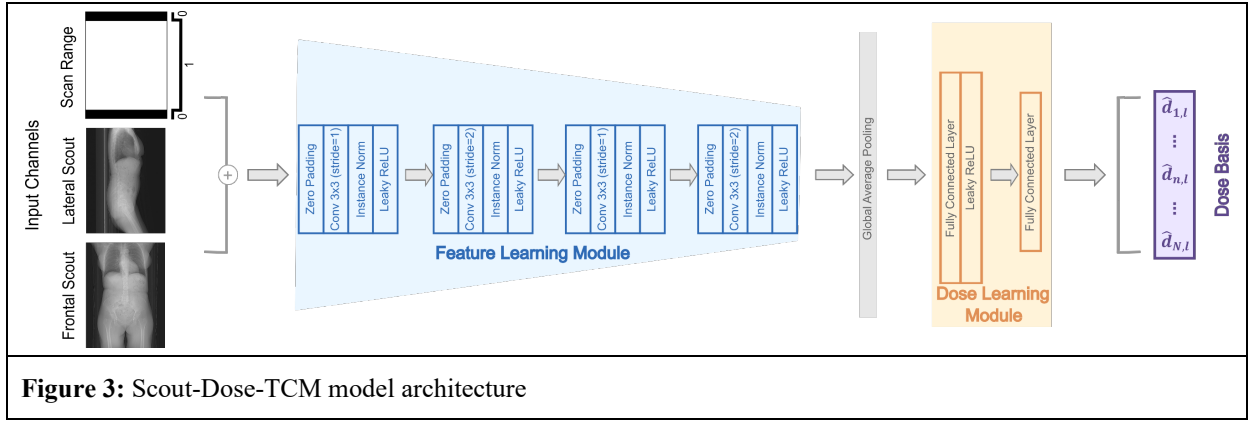

**Figure 3:** Scout-Dose-TCM model architecture

To enforce the DCT basis representation of TCM maps in our network, our model was trained following a customized joint loss function consisting of a basis dose loss and a TCM dose loss. Our basis dose loss accounts for errors in predicting the direct organ-level doses ($d_{l,n}$) corresponding to basis TCM maps ($B_n$), and our TCM dose loss accounts for errors in utilizing our predicted basis organ-level doses to estimate the final organ-level dose ($D_{l,k}$) for a tube current modulation map $TCM_K$. Our final Scout-Net-TCM model loss can hence be defined as

$$Loss = \frac{1}{M}\sum_{1}^{M}\sum_{1}^{L}\left(\sqrt{\frac{1}{n}\sum_{1}^{N}(d_n - \hat{d}_n)^2}\right) + \alpha\left(\sqrt{\frac{1}{k}\sum_{1}^{K}(D_{l,k} - \widehat{D}_{l,k})^2}\right), \quad \text{Eq. 5}$$

where $M$ denotes the mini batch size, $d_n$ the basis dose for DCT basis function $n$, and $D_{l,k}$ represents the dose for organ $l$ and TCM $k$ as described in Section 2.2.

To implement our customized loss function, we generated $K = 50$ random TCM maps with their corresponding discrete cosine transform coefficients $\beta_{k,n}$ satisfying constraints in feasible tube current for each patient. An ablation study was then performed to select the best weighting coefficient $\alpha$ that would balance contributions from basis and TCM dose loss to our final network loss. The final Scout-Dose-TCM model was trained using an Adam optimizer with a learning rate of $2\times10^{-4}$ and a weight decay of $1\times10^{-6}$. Training was performed with a batch size of 8 for up to 100 epochs with the best model selected based on validation accuracy. All models were implemented in Python with the PyTorch framework and run on an Intel Core i9 128 GiB machine with an NVIDIA Titan RTX 24GB.

*Conventional $CTDI_{vol}$-Based Methods*

Following the exponential relationship described in AAPM TG 204 and 220 (Boone *et al.*, 2011; McCollough *et al.*, 2014), the normalized organ dose can be derived from the water-equivalent diameter by

$$Normalized\ organ\ dose = \frac{Organ\ dose}{CTDI_{vol,Global}} = A \times e^{-B\times D_w}, \quad \text{Eq. 3}$$

where $CTDI_{vol,Global}$ and $D_w$ denote the scanner-reported CTDI volume and water-equivalent diameter, and $A$ and $B$ are organ-specific constants. To apply Khatonabadi et al.'s method for accounting for organ dose variations due to non-uniform TCM (Khatonabadi *et al.*, 2013), we computed the organ-specific $CTDI_{vol}$ ($CTDI_{vol,Organ}$) to normalize the organ dose in Eq. 3 using the following:

$$CTDI_{vol,Organ} = CTDI_{vol,Global} \times \frac{I_{Organ}}{I_{Global}}. \qquad \text{Eq. 4}$$

Here $I_{Organ}$ and $I_{Global}$ denote the average organ-level and patient-level tube currents, respectively. To calculate $I_{Organ}$, we leveraged the organ-level segmentation from TotalSegmentator, used for our benchmark organ-level dose calculations, to define the slice locations and corresponding average tube current value on the anatomical section corresponding to the desired organ of interest. The $CTDI_{vol}$ used for our calculations and model fitting was based on the scanner-reported volumetric computed tomography dose index measured using a 32-cm body phantom. Additionally, the water-equivalent diameter ($D_w$) was used as an attenuation-based patient size metric, as previously suggested by Khatonabadi et al. From this point forward, the implementation of Equation 3 normalized by $CTDI_{vol,Global}$ and $CTDI_{vol,Organ}$, will be referred to as the Global and Organ $CTDI_{vol}$ methods, respectively.

*2.4. Performance Evaluation*

A 5-fold cross-validation scheme was used to evaluate the performance of the three prospective organ-level dose estimation methods on out-of-sample cases. For each method, the dataset was randomly divided into five non-overlapping, equally sized folds. In each run, four folds were used for training and one for testing. The Scout-Dose-TCM training set was further partitioned into 80% training and 20% validation subsets for hyperparameter tuning. The mean absolute percent dose error between the prospectively and retrospectively calculated organ-level doses were computed and reported for each patient in the test set across all five folds. Statistical analysis was conducted to evaluate the distribution of organ-level absolute percent dose errors for the three prospective methods across all test patients, including minimum, maximum, and outlier values. Scatter plots were also generated to directly compare the difference between prospectively estimated organ-level doses and their corresponding retrospectively calculated ground truth values.

To assess the statistical significance of our results, we reported the 95% confidence interval (CI) for each organ-level dose error distribution. Because the data were not normally distributed, as determined by the Shapiro-Wilk test, a Wilcoxon signed-rank test was used to evaluate whether differences in estimated absolute percent dose errors between our method and alternative prospective dose estimation methods were statistically significant. A *p*-value less than 0.05 was indicative of a statistically significant difference in this study.

## 3. RESULTS

*3.1. Implementation of Prospective Organ-Level Dose Estimation Methods*

Figure 4 presents representative logarithmic regression fits for the Global and Organ $CTDI_{vol}$ methods across all six organs in a selected cross-validation fold. As shown, organ-level doses exhibited stronger correlation with water-equivalent diameter when normalized by $CTDI_{vol,\,Organ}$ compared to $CTDI_{vol,\,Global}$, except for the bladder, where organ-specific TCM normalization did not yield improved correlation. This figure also illustrates that most patients in our cohort had water-equivalent diameters between 21 and 35 cm and that the average $CTDI_{vol}$ increased with patient size, as expected in CT exams with TCM.

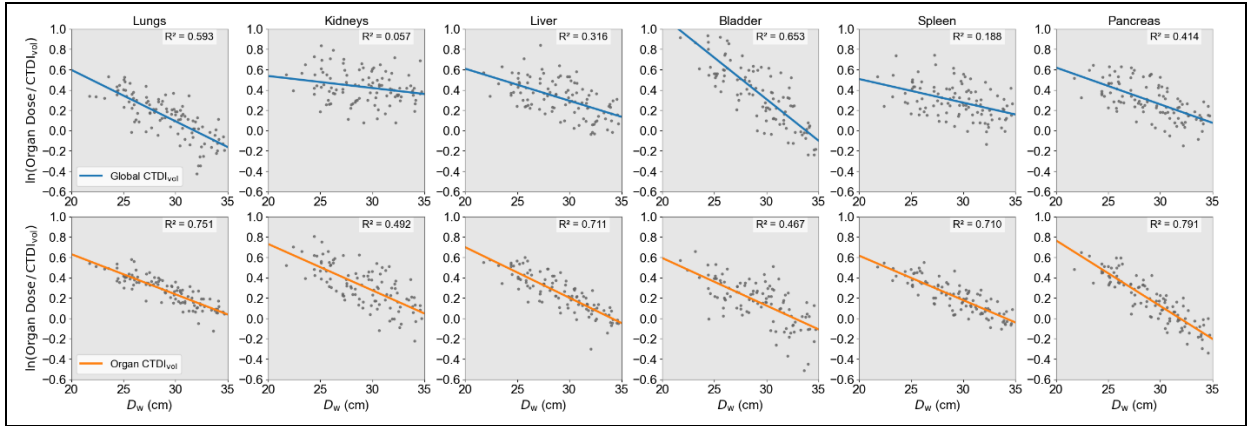

**Figure 4:** Global and Organ $CTDI_{vol}$ methods for lungs, kidneys, liver, bladder, spleen, and pancreas dose prediction in one of the five cross-validation folds.

To evaluate the impact of our proposed joint loss function and identify the optimal α for our Scout-Dose-TCM, we conducted an ablation study using α values of {0, 0.03, 0.05, 0.1, 0.2, 0.5, 0.7, 1, and ∞}. These values were selected to reflect progression from minimal to dominant weighting of the TCM dose loss term, with α = ∞ corresponding to training with only the basis dose loss term. For each α, we computed the relative percent error in organ-level TCM dose prediction on the validation for a set of fifty random, out-of-sample TCM maps not included in our training set. As shown in Figure 5, model performance improved with increasing α up to 0.2, after which further increase in α led to degrading model performance. Based on these results, we selected α = 0.2 as the optimal value for this study.

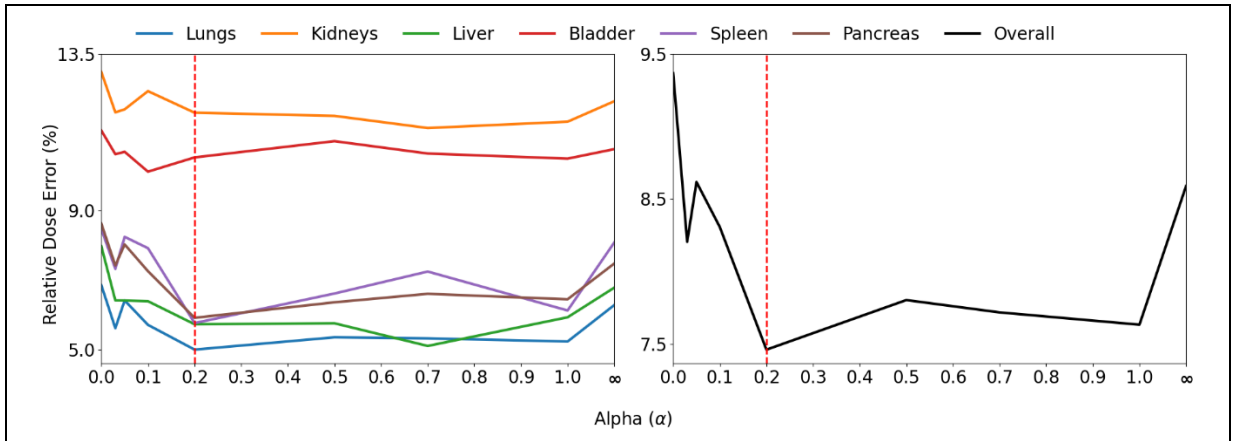

**Figure 5:** Relative percent dose error results for varying joint loss function α values. Left plot shows organ-level errors, and right plot shows the corresponding overall average across validation patient cases and fifty randomly generated out-of-sample TCM maps.

## 3.3. Prospective Dose Estimation Error Analysis

Table II summarizes the distribution of mean absolute percent dose error across the six evaluated organs, including the mean, standard deviation, minimum, and maximum values. These results quantify the discrepancy between the benchmark Monte Carlo-based dose estimates and prospectively predicted organ-level doses based on the patient's original TCM map. Figure 6 presents box-and-whisker plots with 95% confidence intervals, further illustrating the distribution of absolute percent dose error across the three evaluated dose estimation methods. As shown in Table II and Figure 6, Scout-Dose-TCM consistently achieved lower mean and standard deviation of organ-level dose errors compared to the Global and Organ $CTDI_{vol}$ methods. Among the evaluated organs, the lungs exhibited the lowest mean absolute percent dose error across all methods with mean errors of 10.5% (Global $CTDI_{vol}$), 5.5% (Organ $CTDI_{vol}$), and 4.6% (Scout-Dose-TCM). In contrast, the bladder showed the largest discrepancy with values of 15.3%, 13.1%, and 8.89% for Global $CTDI_{vol}$, Organ $CTDI_{vol}$ and Scout-Dose-TCM, respectively. Statistical comparisons using the Wilcoxon signed-rank test revealed significantly lower organ-level dose error from Scout-Net-TCM compared to Global $CTDI_{vol}$ for the lungs, liver, bladder, spleen, and pancreas with all $p < 1 \times 10^{-7}$. Comparisons between Scout-Dose-TCM and Organ $CTDI_{vol}$ also showed significant improvements for liver, bladder, and pancreas with $p$ = 0.005, 0.002, 0.0006, respectively. No statistically significant differences were observed for the kidneys between Global $CTDI_{vol}$ and the other two alternative methods, as indicated by the overlapping 95% confidence intervals in Figure 6.

**Table II:** Summary of absolute percentage dose errors and $R^2$ values comparing three evaluated methods for organ-level dose predictions with reference Monte Carlo dose calculations.

|  | Global $CTDI_{vol}$ | | | | Organ $CTDI_{vol}$ | | | | Scout-Net-TCM | | | |
|---|---|---|---|---|---|---|---|---|---|---|---|---|
|  | Mean±Std | Min | Max | $R^2$ | Mean±Std | Min | Max | $R^2$ | Mean±Std | Min | Max | $R^2$ |
| **Lungs** | 10.5±9.20 | 0.29 | 53.2 | 0.57 | 5.48±4.80 | 0.06 | 26.1 | 0.86 | **4.63±4.31**[a] | 0 | 20.6 | 0.89 |
| **Kidneys** | 12.4±9.76 | 0.01 | 44.3 | 0.73 | 12.6±8.71 | 0.16 | 47.0 | 0.69 | **10.7±7.36** | 0.08 | 30.1 | 0.78 |
| **Liver** | 11.9±8.52 | 0.06 | 37.0 | 0.66 | 7.45±7.07 | 0.04 | 51.7 | 0.83 | **5.24±4.22**[a,b] | 0.05 | 23.2 | 0.92 |
| **Bladder** | 15.3±10.9 | 0.19 | 49.7 | 0.26 | 13.1±11.7 | 0.09 | 64.0 | 0.48 | **8.89±6.99**[a,b] | 0.01 | 33.9 | 0.68 |
| **Spleen** | 13.4±9.44 | 0.03 | 49.3 | 0.65 | 6.92±5.49 | 0.18 | 21.7 | 0.89 | **6.16±5.14**[a] | 0.06 | 27.6 | 0.91 |
| **Pancreas** | 11.2±8.16 | 0.18 | 40.7 | 0.69 | 8.40±6.59 | 0.02 | 31.4 | 0.79 | **5.71±4.59**[a,b] | 0 | 28.3 | 0.90 |

[a] Statistically significant difference to Global $CTDI_{vol}$ ($p < 0.05$)
[b] Statistically significant difference to Organ $CTDI_{vol}$ ($p < 0.05$)

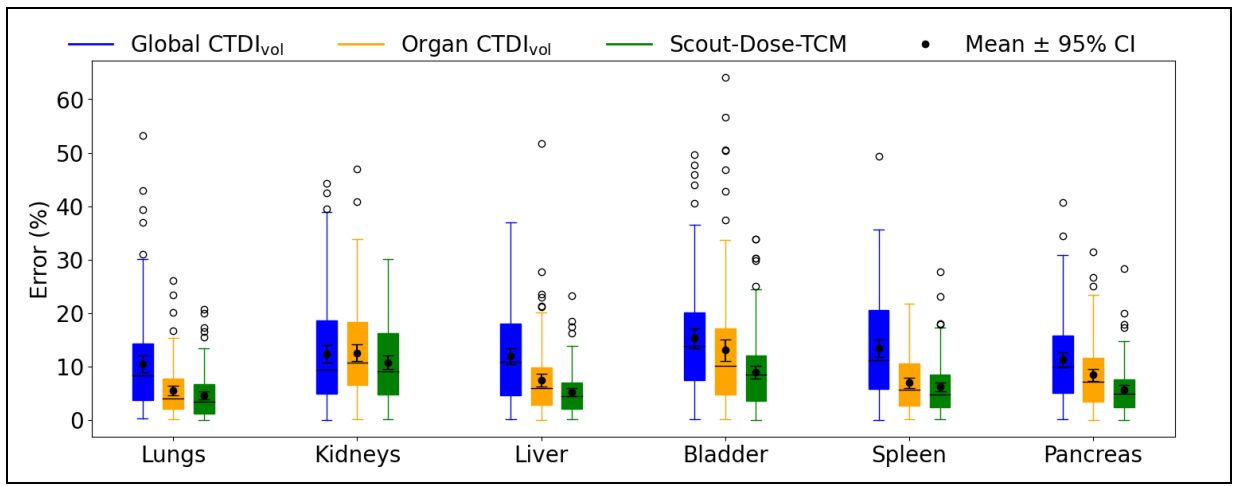

**Figure 6:** Distribution of absolute percent dose errors between final dose predictions from three evaluated methods and reference Monte Carlo dose calculations, analyzed across 130 patients and six studied organs.

Figure 7 shows scatter plots comparing predicted organ-level doses against benchmark values for the six evaluated organs and the three prospective dose estimation methods. The solid black identity line represents perfect agreement between predicted and ground truth values. For each method, shaded regions enclose the central 95% of data points for each method, and corresponding $R^2$ values (coefficient of determination) quantify the strength of correlation between predicted and benchmark values. Across all six organs, Scout-Dose-TCM consistently yielded higher $R^2$ values than Organ and Global $CTDI_{vol}$ methods, showing stronger agreement with benchmark values. Furthermore, the scatter distributions for Scout-Net-TCM were more tightly clustered around the identity line, with narrower 95% confidence regions, further showing the improved predictive performance of the proposed Scout-Dose-TCM over the alternative methods.

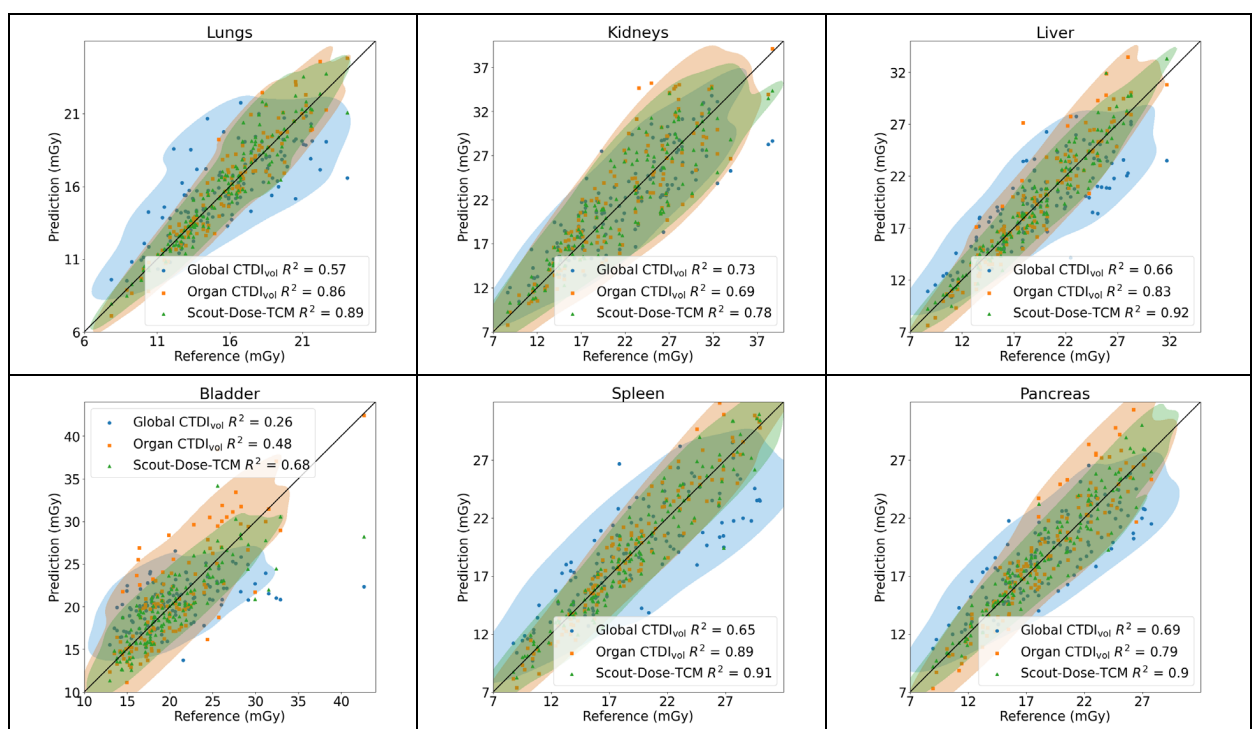

**Figure 7:** Scatter plots showing the distribution of reference and predicted organ-level doses for six organs through three investigated methods. Shaded regions denote the distribution containing 95% of the data for each method. The straight line denotes the identity line corresponding to perfect correspondence between reference and predicted doses.

## 4. DISCUSSION

In this study, we proposed a deep-learning based model for prospective organ-level dose prediction using scout images from tube current modulated CT exams. Unlike previously proposed methods, Scout-Dose-TCM leverages anatomical, size, and attenuation information encoded in the scout images, as well as the periodic structure of TCM profiles, to directly estimate organ-level doses for any feasible TCM pattern. To assess performance, we compared our model to the AAPM TG

204 method and to the organ-level TCM-adjusted version proposed by Khatonabadi et al. Scout-Dose-TCM achieved an organ-level mean absolute relative dose error of 7%, outperforming the Organ and Global CTDI$_{vol}$ methods which yielded 9% and 13% errors, respectively. In addition, Scout-Dose-TCM produced a more compact error distribution. For context, organ dose errors of 7% have been attributed to imperfect organ segmentation (Schmidt *et al.*, 2016), up to 10% when measured in physical phantoms with thermoluminescent dosimeters (TLDs) (Principi *et al.*, 2021), 15% when comparing MC simulation against TLD measurements (Kalender *et al.*, 2014), and 20% when using convolution-based estimation (Tian *et al.*, 2016). In addition to improved accuracy, Scout-Dose-TCM enabled real-time dose estimation, generating organ-level dose predictions in 0.006 seconds per case. In contrast, retrospective MC-GPU dose calculations require 1-2 hours to generate full 3D voxel-level dose maps. While Monte Carlo remains the gold standard for detailed dose assessment, our method offers fast, organ-level dose estimates suitable for prospective planning and TCM optimization in a fraction of the time.

Scout-Dose-TCM also notably improved bladder dose prediction, an organ that has been previously challenging to estimate. This improvement may be attributed to Scout-Dose-TCM's ability to infer spatial context which enables it to capture the bladder's anatomical coverage and its location within the pelvic region, which is surrounded by high-attenuation bone and subject to sharp modulation gradients. Although Khatonabadi et al. did not evaluate the bladder in their original study, Tian et al. reported bladder dose errors of 12% and 22% when modeling TCM profiles at four different modulation strengths using a convolution-based dose-field model. The highest error occurred at the strongest modulation setting tested (Tian *et al.*, 2016). The improved performance in our model suggests that Scout-Dose-TCM's feature-learning module was able to effectively encode contextual size information from scout images and learn spatially varying TCM behaviors particularly in highly variable anatomical regions such as the pelvis. In contrast, kidney dose prediction proved to be challenging across all three methods. This may be attributed to asymmetries between the left and right kidney in terms of size (typically differing by 10-15 mL), z-positioning (with the right kidney generally located lower than the left), and surrounding tissue composition that is not fully captured by water-equivalent diameter or scout-based estimations (Janoff *et al.*, 2004; Cheong *et al.*, 2007; Summerlin *et al.*, 2008; Tanriover *et al.*, 2015). Additionally, kidneys are small relative to other solid organs in the abdomen, making them particularly sensitive to segmentation inaccuracies (Heller *et al.*, 2021). Respiratory motion may also introduce variability in kidney positioning between scouts and diagnostic acquisition, which could further degrade the accuracy of dose predictions. Although Khatonabadi et al. and Tian et al. did not report challenges in estimating kidney dose under non-uniform TCM, both studies included relatively small patient cohorts (n = 39 and n = 60, respectively), (Khatonabadi *et al.*, 2013; Tian *et al.*, 2016) potentially limiting the visibility of such limitations. Future work will focus on improving kidney dose estimation by incorporating 3D spatial dose distributions from full-resolution or deep learning-generated CT images, accounting for respiratory motion, or by estimating dose separately for the left and right kidneys.

Our study also offered new insights into classical prospective organ-level dose estimation methods. By re-implementing the Global and Organ CTDI$_{vol}$ approaches on a larger patient cohort using modern open-source segmentation tools, we confirmed the potential of Organ CTDI$_{vol}$ to reduce dose prediction errors (by 4%) compared to Global CTDI$_{vol}$. Notably, we observed that Organ CTDI$_{vol}$ normalization meaningfully improves dose estimation for the lungs and spleen, contrary to prior findings. Our re-implementation also highlighted the limitation of the Organ CTDI$_{vol}$ method, particularly for organs like the bladder. Minimal correlation improvement was observed between water-equivalent diameter and organ dose for bladder when normalized by $CTDI_{vol,organ}$ in comparison to $CTDI_{vol,global}$. This may reflect challenges in capturing variability in the bladder's anatomical coverage through a single, average size metric, as well as potential artifacts (e.g., metal artifact from hip implants) which may affect organ attenuation and calculated dose values. Future iterations of the Organ $CTDI_{vol}$ method could explore region-specific water-equivalent diameter metrics, an idea proposed by previous investigators (Khatonabadi *et al.*, 2013), or pivot towards deep learning-based approaches such as Scout-Dose-TCM, which may be inherently better suited at capturing organ-specific anatomical coverage. Additionally, future studies should examine the impact of imaging artifacts on organ dose estimation for both traditional and deep learning–based methods. Finally, while Organ $CTDI_{vol}$ relies on manually defined organ segmentations from retrospective data, true prospective implementation would require scout-based organ segmentation. Our group previously developed deep learning methods for automatic organ segmentation from scout images and demonstrated the potential of these segmentations in further refining organ-level dose predictions from our Scout-Dose model (Medrano *et al.*, 2023; Fang *et al.*, 2025). Future research could explore combining the strengths of both approaches: integrating deep-learning–based segmentation for prospective organ definition with the interpretability of fitting-based dose estimation models.

While this work constitutes an encouraging proof-of-concept for scout-based, prospective organ-level dose estimation, several limitations need to be addressed before clinical implementation. Our model was trained and tested on contrast-enhanced chest-abdomen pelvis scans at 120 kVp with contrast simulated as dense water in Monte Carlo simulations. Organ dose has been reported to increase by up to 50% in the presence of contrast media (Sahbaee *et al.*, 2017; Mazloumi *et al.*, 2021). As part of our future work, we will train specialized Scout-Dose-TCM models for different scanning protocols to enable protocol-specific organ-dose estimates. Additionally, our method requires proprietary vendor information for accurate Monte Carlo dose estimates. While this limitation applies broadly to Monte Carlo dosimetry, previous studies have demonstrated that scanner-specific parameters such as TCM, bowtie filter geometry, and anode heel effect can be approximated using ray-tracing algorithms or calibration methods (Keat N, 2005; Whiting *et al.*, 2014; Ming *et al.*, 2017). Future efforts could evaluate whether such techniques can extend Scout-Dose-TCM's applicability across scanner platforms. Furthermore, in this study, we used frontal and lateral scout views as inputs to Scout-Dose-TCM. This design choice was informed by our prior findings with Scout-Net, where incorporating two views improved organ dose prediction accuracy under uniform TCM by up to 40% (Imran *et al.*, 2021, 2022, 2023). Given that non-uniform TCM introduces additional complexity in dose estimation, a two-view approach provided a robust starting point. Future work may explore the feasibility and trade-offs of using a single scout view to reduce patient exposure and simplify acquisition protocols. Given that our patient cohort was drawn from a single institution, the water-equivalent diameter in our data set ranged of 20–35 cm; expanding this range to include pediatric and obese patients constitutes an important next step. Furthermore, while Khatonabadi et al. suggested separate fits for male and female patients, Scout-Dose-TCM's strong performance suggests that a unified model may suffice. More generally, future studies could explore larger, more diverse datasets. Furthermore, our study was limited to six organs to enable focused, organ-specific analysis and facilitate direct comparisons with prior studies. By prioritizing depth over breadth, we focused in this study on identifying challenges in prospective scout-based organ dose prediction, such as the impact of anatomical variability and TCM. Future work will expand this analysis to additional organs (such as those needed to calculate effective dose) and explore direct effective dose prediction to broaden the clinical utility of Scout-Dose-TCM.

Overall, our proposed Scout-Dose-TCM framework was able to perform prospective, real-time, and patient-specific organ-level dose estimation from scout images of tube current-modulated CT exams. Beyond supporting personalized radiation safety assessment, prospective organ-level dose estimation opens new possibilities for task-specific TCM optimization in clinical practice. A promising future direction will be integrating Scout-Dose-TCM into task-based TCM optimization frameworks, laying the foundation for personalized CT protocols that balance diagnostic image quality and radiation dose (Klein *et al.*, 2022; Wang *et al.*, 2024). Through this work, we highlight the potential of combining prior knowledge in CT dosimetry with modern deep learning techniques to enable real-time, patient-specific dose estimation and pave the way for more personalized CT imaging.

## 5. CONCLUSIONS

Herein, we have proposed Scout-Dose-TCM, a deep learning–based model for prospective, patient-specific organ-level dose estimation in tube current–modulated CT exams. By leveraging the anatomical and attenuation information encoded in scout images, Scout-Dose-TCM predicts organ doses without requiring post-acquisition reconstruction and manual segmentation. Our comparative study demonstrated that Scout-Dose-TCM resulted in improved dose prediction accuracy (7%) compared to traditional Global and Organ $CTDI_{vol}$ methods (13% and 9%, respectively), particularly for anatomically complex regions such as the bladder. While further refinements are needed for widespread clinical application, this work establishes the feasibility of scout-based organ dose estimation under non-uniform TCM and highlights the potential of deep learning models to advance patient-specific CT dosimetry. Future work will focus on expanding Scout-Dose-TCM to diverse patient populations, incorporating additional organs, and integrating this approach into task-based TCM optimization frameworks. By enabling prospective, real-time, patient-specific organ-level dose estimation, Scout-Dose-TCM represents a promising step toward the development of more personalized CT protocols that balance radiation dose and diagnostic image quality.

## ACKNOWLEDGMENTS

This work was supported by GE HealthCare and Stanford Propel Postdoctoral Fellowship.

We thank Jooho Lee for his valuable input and thoughtful suggestions during the development of this work.